# Solar Synergy: Innovative Strategies for Data Centers Energy Efficiency and Sustainability

Alnawar J. Mohammed, and Qutaiba I. Ali, *Member, IEEE*

*Abstract*— One of the current trends related to data centers is providing it with renewable energy sources. This paper suggests an analysis technique for a model uses solar panels energy to power a data center consists of 100 traditional servers, physical infrastructures, 5 backup batteries. The analysis passes through three phases: Initially, the power consumption model of the data center is proposed to show the variation in traffic and total energy consumed. Then, a solar system model is designed according to the power needed. The suitable size of PV was found to be 613 solar panels that delivers an energy of 2.967MWh. At the last phase, the desired battery capacity is chosen as (10000Ah×48V) to accommodate the solar energy production. The problem occurred when there is a shortage in solar batteries' energy especially, during the night when no sunlight exists. An integration method between using the solar energy with traffic consolidation technique (ON/OFF algorithm) is suggested to overcome this issue and extend the battery life. Finally, this model is evaluated by the number of active servers according to the number of batteries available since it found that it reduced to (10-35) % of the total number when only 4 batteries are used and the effective power provided will be between (1.68-1.904) MWh with an energy saving of about 867KWh for the case of only 10% of the total servers are active.

## I. INTRODUCTION

Most of the current trends that emerge in the energy consumption field have recently revealed the need for sustainability in Information and Technology (ICT). The environmental problem of greenhouse gases effect through carbon dioxide emission is well understood. The term green data centers were brought in the modern ICT to support the green computing which is the efficient design, use, and integration of ICT resources, utilizing minimum amount of energy for reducing the environmental affect [1]. Different promising strategies aimed at diminishing greenhouse gas (GHG) emissions and associated pollution involves enhancing energy efficiency. This can be accomplished through virtualization, load balancing and consolidation and so on. Such a combination of techniques requires making the resources of data centers available to customers, while also distributing these resources optimally across the minimal physical infrastructure and powering down idle physical machines. An additional method, known as Dynamic Voltage and Frequency Scaling (DVFS), dynamically adjusts the frequency and voltage of processors operating servers. It conserves energy when machines are idle or running fewer demanding servers; however, it still complies with the Service Level Agreement (SLA) [2].

The power supply form has a huge effect on carbon footprint emission where, some sources are with high carbon emitting like coal, on the other hand, wind and sun as renewable energy sources are lower carbon emission. Thus, introducing the proactive measures of sustainability and efficiency and investigate how to use the renewable energy power to drive the data center efficiently [3]. Solar energy production experiences changes during different seasons of the year, and it is higher during summer, and lower during winter. The fact is that this energy can be charged by generation from renewable energy sources if that generation is available and can discharge that energy when required again. For instance, short-term storage is capable of powering a few hours [4]. Solar energy can be used in different application like designing Road Side Unit (RSU), which is a critical component of VANET with a solar energy harvester for a renewable power supply. For the power supply, an interface in the form of a power management module is provided for the RSU to operate independently from the standard power sources. This implies that the design process selects appropriate subcomponent such as the solar panels and batteries; and at the same time integrates with CDC technology to manage power usage [5], [6].

**Problem Statement:** The main issue of this research is to reduce energy consumption of ICTs in data center and thus reduce the greenhouse gases effect through carbon dioxide emission by powering the data centers with clean solar energy which is varies on daily and seasonally basis. This variation can cause a shortage in battery life. For increased power saving and preserve battery life a power management technology must be emerged with solar energy system.

**Related Work:** E. Sheme et al. [7], discusses the possibility of satisfying the energy needs of data centers located in high latitude countries such as Finland by the use of renewable energy sources like wind and solar power. It examines the relationship between the number of wind turbines and the required size of solar panels for reaching the necessary level of renewable energy generation. Five cases are provided, which include different levels of wind and solar installment, and all of them prove that the hybrid system is more beneficial because it provides a maximum of hours with surplus electricity in a year. The results indicated, that although adding wind and solar energy sources could increase surplus hours, the latter could decrease if over-reliance on solar panels because of it depends on sunlight.

M. Yousefi and P. Ghalebani [8] discuss the significance of electrical energy supply in data centers noting the general rise for energy consumption due to advancement in IT and data center businesses. It emphasizes the importance of using renewable energy forms especially solar energy for the reduction of using the conventional energy sources such as fossil energy. This research addresses the possibility of incorporating solar energy systems with the conventional power systems in order to provide continuous power supply for data centers.

**Contributions:** the contribution of this paper is stated below:
1) Analyze the power and energy consumption of a date center of 100 server, the required solar panels according to power load, and the suitable number of batteries for storage.
2) Providing an integration between two energy efficient Technologies: the solar cell model to provide renewable energy source for the data center and the ON/OFF switching for put the idle servers on OFF state to extend the battery life.
3) Evaluating the number of active servers and the variation of power consumed by the data center according to the available number of batteries.

The rest of this paper is divided into four sections. Section II presents background and mentions different power management technologies. The methodology for energy efficient model, solar energy system, and the battery storage model is analyzed in Section III. according to the energy consumption model in data centers and the solar energy System, Section IV shows the results and discuss the results. Finally, conclusions and future works are presented in Section V.

## II. BACKGROUND

This section gives a brief description of energy consumptions and power management techniques for the entire data center. In the following, we have presented some state of art works regarding energy efficiency in data centers.

### A. Power Management Techniques

The energy consumed at a particular data center is dependent on its properties of the hardware, networking layout, storage layout, and many more attributes. Energy optimization can be achieved through simulation, resource utilization, auto scaling of infrastructures, energy-based scheduling, disk space management, VMotion (migration of live virtual machines), low carbon emission and many others. All of these techniques can be implemented at relevant component level to achieve the goals of green computing [9]. On the other hand, software level modelling activity is significant for energy efficiency since the software components consume power to execute tasks. The software development should be able to draw some benefits from achievements at the hardware level. If the software generated is not as efficient as hardware, technical advances require a large number of resources. The overall energy consumption will remain high, thereby defeating the idea of constructing green data center [10]. This situation implicates that refining the efficiency of infrastructure and incorporating sustainable energy resource can contribute a lot to the achievement of the energy efficiency target [11]. In this paper, an ON/OFF power management technology is utilized to place the idle devices in a power off state. Table I. summarizes power management technologies in data centers that uses traffic consolidation and switch OFF unnecessary devices.

### B. Renewable Energy Techniques

Most types of renewable energy, like solar energy, wind, water, biomass, and hydrogen provide a significant volume of energy consumed in data centers. The incorporation of renewable energy resources in existing power systems can be critical in reductions in carbon outputs. Out of these sources, wind and solar have been the favorite among the researchers for their cheap cost and environmentally friendly nature. But these sources are very unpredictable. Thus, there is a need for a battery to store the excess energy so that it can feed the load as needed [11], [12]. Not only data centers, renewable energy sources can also be deployed to improve communication and data transfer in vehicular networks, such as by integrating the proposed system (ECHIDS) which provides security and the use of solar energy for power management [13],[14]. Such clean energy solutions have some constraints. First, the initial investment costs and the costs of deploying such service is relatively expensive. Second, those sources rely on weather because the solar operates during sunny days. Additionally, their locations are also playing a great role in reducing the impact of such solutions [15]. This has eased the ways through which data centers can cut down on fossil fuel usage and lower their emissions of carbon. Major tech giants, have begun to integrate renewable energy sources into their data center operations [16].

- Apple's Commitment to Clean Energy: Apple announced that all data centers, are powered by 100% renewable energy. using solar, wind, and other renewable sources.
- Amazon Web Services (AWS) Wind Farms: for its large network of data centers, AWS announced its plans to harness power from sources such as wind farms.

## III. RESEARCH METHODOLOGY

In this section, the proposed model depends on using renewable energy (solar panels) to power a data center of 100 traditional server with its auxiliary networking and other physical infrastructure like cooling and power system. In addition to use an ON/OFF switching technology to switch OFF the idle servers for energy saving in case of there is a shortage in power delivered by the batteries especially at night. Fig. 1, illustrates a structure for the internal infrastructure of the data center. The analysis of this model passes through three phases: Data Center Power Consumption Model, Solar Energy System, and Battery and storage Model.

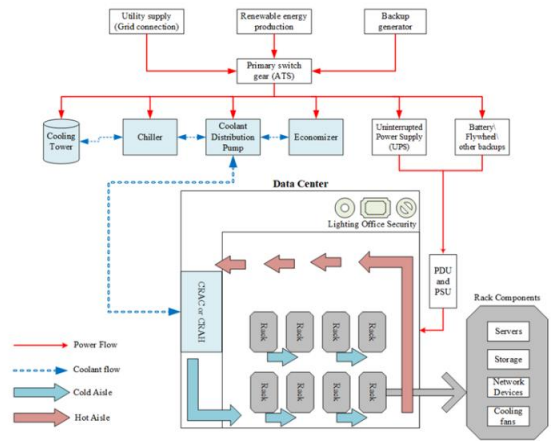

Fig. 1. The internal infrastructure of the data center

TABLE I. POWER MANAGEMENT TECHNIQUES

| Ref. | Technology used | Environment | Objective | Tools | Evaluation |
|---|---|---|---|---|---|
| [17] X. Wang and et.al, 2015 | • Traffic Consolidation | Traditional Data Centers | • Efficient traffic streams onto a few links and switches in a DCN and power off underutilized network equipment to save energy. | • Simulation of<br>• two baseline design: ElasticTree and Google | • Energy Saving |
| [18] L. Zhou, 2019 | • Traffic Consolidating | Data Center with SDN Design | • Combining traffic and shutting unnecessary network interfaces to conserve power in the data center. | • Testbed<br>• 8 blade servers and 6 switches<br>• 2-layer leaf spine DCN topology (Real traces) | • Energy Saving<br>• Packet Drop<br>• Application Latency |
| [19] Y. Liu and et. al., 2020 | • Traffic Aware Technology | Traditional Data Center | • Traffic is employed in order to forecast energy consumption.<br>• Examine how the PUE and the latitude of data centers can enhance the Australasian power effectiveness. | • Romonet's Global Site Analysis Tool (SAT) | • The annual<br>• The annual data center usage<br>• PUE and Latitude relationship<br>• Energy consumption and $CO_2$ emissions |
| [20] L. LIN and et. al., 2020 | • Traffic Aware Technology | Cloud Data Centers | • Save energy by preventing the use of as many active physical machines (PMs) as possible to support VM placement to allow the use of fewer waking sleeping PMs. | • Extensive simulations with FatTree Topology of the DC | • Network Cost<br>• Energy Consumption |
| [21] N. Hogade and et. al., 2022 | • Traffic Aware Management | Geo-Distributed Heterogenous Cloud Data Center | • Reduces energy costs and cost of inter-data center data transfer.<br>• The problem of workload distribution is stated as a non-cooperative game and a game theoretic approach is proposed for the intelligent dynamic workload partition and allocation to data centers over time. | • Simulation | • Energy consumption<br>• Network costs<br>• Queueing delays |
| [22] Lazim Qaddoori S, Ali Q, 2023 | • Machine Learning (ML) and data aggregation | Data aggregator Advanced Metering Infrastructure (AMI) (HAN, NAN and WAN) | • lightweight architecture for a smart meter that utilizes machine learning algorithms for detecting abnormal power consumption. | • Machine Learning (ML) Models | • Effectiveness and efficiency of the model |

**Phase. 1: Data Center Power Consumption Model**

The energy consumed by a data center can be divided into two parts: energy used by IT equipment (servers, networks,.) and that are used by non-IT equipment facilities (e.g., cooling system, other infrastructures), see Fig. 2.

Servers are distinguished as the primary IT device that use significant amount of energy. The power consumption model of the server in (Equ. 1) assumes that the relationship between server power consumption and server CPU usage, is proportional [23]. On average of idle server state takes slightly large power of approximately 70% of the power of the operational active server which work with the maximum CPU load [24].

$$P(U)_{Server} = K * P_{max} + (1 - K) * P_{max} * U \quad \text{-------1}$$

Where,
Pmax: is the maximum power consumed when the server is fully utilized;
K: is the fraction of power consumed by the idle server (i.e. 70%).
U: is the CPU utilization of the server.

Also, it can be written as in (Equ.2):

$$P(U)_{Server} = P_{idle} + (P_{max} - P_{idle}) * U \quad \text{--------- 2}$$

The energy consumption of all the servers $P_{Servers}$ can be defined in (Equ.3) as follow:

$$P_{Servers} = \sum_{i=1}^{N} P(U)_{Server(i)} \quad \text{--------3}$$

Where, i = 1,2,3 …..,N, means the sequence of servers,
N: the maximum number of servers in data center

**Steps to Compute the Total Power Consumption of the Data Center**

1- Find the Server Power Consumption.
$P(U)_{Server} = K * P_{max} + (1 - K) * P_{max} * U$
U: Server CPU Utilization
K: Fraction of power consumed by the idle server (i.e. 70%)
$P_{max}$: maximum power consumed by a server

2- Determine the power Consumption of All Servers in Data Center.
$P_{Servers} = \sum_{i=1}^{N} P(U)_{Server(i)}$
i: 1,2,3,4, ….., N
N: maximum number of servers in data center

3- Calculate the IT Load Power Consumption
IT Load Power Consumption = $P_{Servers} + P_{Networking\ Equipment} + P_{Storage\ Equipment}$

4- Determine the Power Consumption of the non_IT equipment (Cooling Units and other Physical Infrastructures)

5- Find the Total Power Consumption of the Date Center
Total Power Consumption = IT Load Power Consumption + non IT equipment Power Consumption

Fig. 2. Procedures for total power consumption in data centers.

On the other hand, non-IT equipment facilities, the cooling and environmental control system, is used to maintain the temperature and humidity of the data center. This mainly contains the Computer Room Air Cooling System (CRAC) unit to ensure the reliable coolant flow in the data hall. The highly-dense IT loads generate enormous amount of heat in data center, which is handled by the cooling load [23].

The power consumed in all servers in addition to the that consumed in the corresponding cooling systems can be estimated by assuming that a 2.5KW cooling unit is used to cool 5 servers, which is in agrees with the assumption of ref. [24] that it takes 1 to 1.5W to cool the servers for every watt of power they consume, given the maximum power of a server is assumed as 250W.

For the networking equipment, storge equipment and other physical infrastructures to be considered in calculations of the total power consumption in data center, which is mainly IT load (the servers, networking equipment, and external storage) and non-IT load (PDUs, UPSs, Cooling System), a measurement tool by Schneider Electric (see Fig. 3) is used by set the maximum number of servers in data center to 100 traditional servers, having 1 to 2 CPU/GPU each.

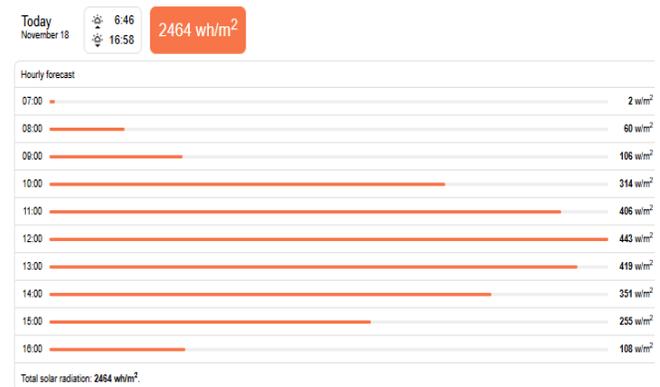

Fig. 3 the input parameters to calculate the total power of data center

The donut chart explains the power consumed by each equipment of the data center this include the IT equipment as 150KW and the physical infrastructure as 60KW. It can be seen that, the total power consumption for the data center is equal to 210KW (see Fig. 4)

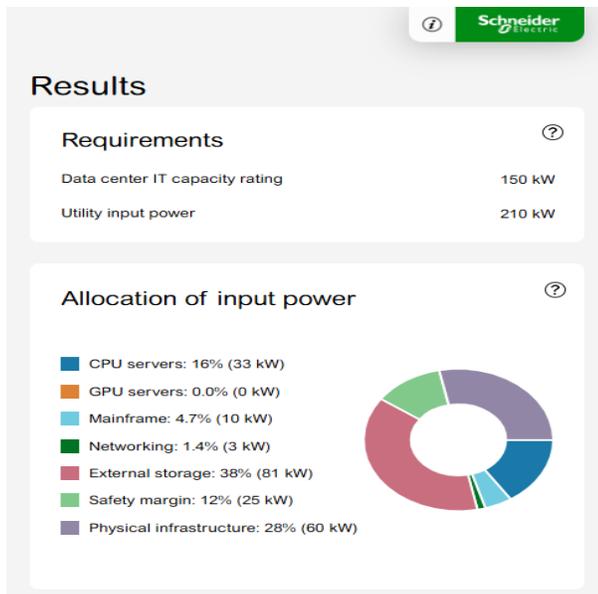

Fig. 4. The total power consumption of the data center

**Phase. 2: Solar Energy System**

Solar power is generated when the sunlight is transformed into power. Solar energy is not available during the night [8]. The radiation graph of sun in Mosul city for one day (see Fig. 5) is taken into account in the analysis to identify the hours of irradiance during the daylight. This graph explains the absorbed power distribution (w/m$^2$) at each hour from 7AM to 5PM, the maximum power absorbed is (443 w/m$^2$) at 12:00PM. From the figure we can see that the power generated from the solar system per 1m$^2$ is equal to 2.464KWh while the total power consumed by data center is 2.94MWh. In this case, the data center needs for 613 solar panel of size (6.5ft × 3.25ft) this means (1.98 m× 0.99 m) each contains 72 solar cell installed on an area of 1200m$^2$ to provide 2.967MWh. There are two important metrics related to renewable energy must be evaluated:

Fig. 5. The radiation graph of sun in Mosul city

*1) Minimum percentage supply*

By combining the findings produced by solar energy production and energy consumption allows constructing the green energy coverage. The coverage described by analyzing the ratio between the renewable energy being produced by 1 m$^2$ solar panel and the data center energy consumption [7], MPS is given by (Equ. 4).

$$MPS = \frac{Renewable\ Energy\ Production\ (KWh)}{Total\ Energy\ Consumption\ (KWh)} \times 100 \quad \text{----- 4}$$

In our case, the renewable energy being produced by 1 m$^2$ solar panel is 2.464KWh and the total power consumed by data center is 2.94 MWh then MPS equals to 8.38%. This means the renewable power produced by 1m$^2$ of the solar panel is 8.3% of the total power.

*2) Renewable energy variation*

As it has been noticed, renewable energy production fluctuates from one period to another due to the prevailing weather conditions. The data center consumption is also variable based on the workload at different time intervals as well. This implies that at one time there is a higher production rate than the consumption rate, at other times it is the reverse, this is given as ΔE in (Equ. 5) [7].

$$\Delta E = Renewable\ Energy - Energy\ Consumption \text{ --- 5}$$

ΔE of our system is 27KWh, where the renewable energy is 2.967MWh and the energy consumed by data center is 2.94 MWh (ΔE meaning that more energy is produced than what is consumed).

**Phase. 3: Battery and Storge Model**

While energy storage is important as a backup source of electricity, it enables more flexible use of renewable power. Energy storage can be charged from renewable generation when it is available and can then discharge that power when it is later needed. Energy can be stored using a range of technologies. In this paper, 5 backup batteries (10000 Ah × 48v) are required with 80% efficiency and a battery capacity of 0.48MWh each. Fig. 6 shows the process of calculating the solar generated energy to be stored in batteries with the number of batteries required.

---

**Steps to Calculate the Number of required Batteries to Power the Data Center for 14 hrs**

1- Find the Total Energy Consumption of Data Center.
   **Total Energy= IT Load + Physical Infrastructure**
2- Specify the backup time.
   In our case the night hours = **14hrs** from (**5PM to 7AM**)
3- Convert the total energy consumption to watt-hr (Wh)
   **Total Energy (Wh)= Total Energy × Backup Time**
4- Calculate Battery Capacity in watt-hr (Wh).
   **Battery Capacity (Wh)= Battery Capacity (Ah)× Battery Voltage (V)**
5- Determine the Effective Energy using Battery Efficiency (80-90%).
   **Effective Energy (Wh)= Total Energy (Wh)× Battery Efficiency**
6- Find the Required Batteries according to effective energy.
   **Required Batteries = Effective Energy (Wh)÷Battery Capacity (Wh)**

Fig. 6. The process of using solar energy system in data center

---

Inverter Selection: Converters or inverters are important devices that transform the current produced by solar panel – direct current (DC) – into the type of current used in data centers – alternating current (AC) (see Fig. 7). The suitable inverters which work well with your solar panels and your battery system must be chosen[25].

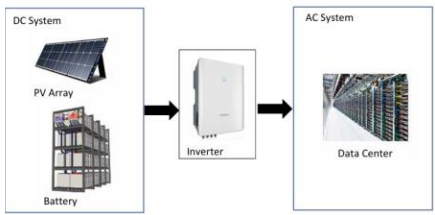

Fig. 7. The DC-AC Convertor

## IV. RESULTS AND DISCUSSIONS

With the provided results of the model being implemented in this paper the evaluation process will be as follows:

*1) Data Center Energy Consumption Results:* for the data center proposed in this paper the power consumption loads is analyzed using Schneider Electric tool which is found to be 0.21 MW. Taken into account the total equipment including IT equipment (100 traditional servers, networking, and storage) and non-IT equipment (cooling units and auxiliary infrastructure). This power load is consumed separately between data center equipment (see Fig. 8). The line graph shows that the main parts that consumes the most of power are servers, storage, and physical infrastructures including cooling units.

*2) Solar Energy System Results:* to find the suitable number of solar the radiation graph of sun in Mosul city is examined, this graph gives an information about the hours of radiation from 7AM to 5PM this means that data center needs to use batteries during the night from 5PM to 7AM. the amount of power (Wh) in each

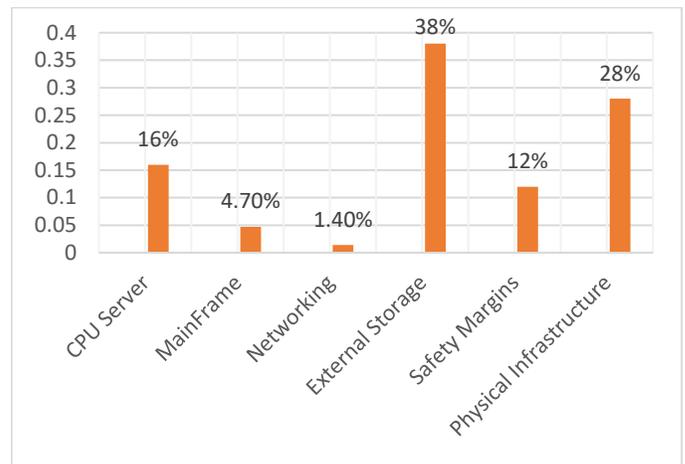

Fig. 8. Power Consumption Percentage

1m2 of the solar panel during the sunlight hours is also given as (2464Wh/m2) where the maximum energy absorption during the daylight was (443 w/m2) at 12:00PM. The total power consumption of data center if all the servers are active gives an indication about the number of solar panels required. Considering (1.98m× 0.99m) solar panels are used. So, the data center needs for 613 solar panel each of 72 solar cells for 2.967MWh production. Then, the green energy coverage metric is examined by MPS and $\Delta E$ which are found as 8.38% and 27KWh respectively.

*3) Battery and Storage System Results:* the calculated value of the data center energy consumption with the corresponding cost like the required number of batteries is determined accordingly. Then, the number of active servers is evaluated in case of using the ON/OFF switching power management technology if there is a shortage in batteries' storage during the night and to extend battery life. Table. II shows the relation between the data center energy consumption, the number of active servers, and number of batteries required. Results shows that if the data center runs only 4 batteries, it must reduce the number of active servers to (10-35) %. While, the effective power provided will be between (1.68 - 1.904) MWh.

TABLE. II The relation between the data center energy consumption, the number of active servers, and number of batteries required.

| No. of Active servers | EVALUATION | | | |
|---|---|---|---|---|
| | Total power consumption in data center (MW) | Total power consumption in data center (MWh) | Effective Energy (MWh) | No. of Batteries |
| 100 | 0.21 | 2.94 | 2.3 | 5 |
| 80 | 0.2 | 2.8 | 2.24 | 5 |
| 60 | 0.18 | 2.52 | 2.016 | 5 |
| 35 | 0.17 | 2.38 | 1.904 | 4 |
| 10 | 0.15 | 2.1 | 1.68 | 4 |

Finally, the power saving is calculated after switching OFF the idle servers (see Fig. 9). Different cases are examined and it is clear that 167KWh out of 2967KWh a total energy provided by the solar system is achieved for the case of 80% of the operating servers, whereas it jumps to 867KWh for the case of 10% of the operating servers.

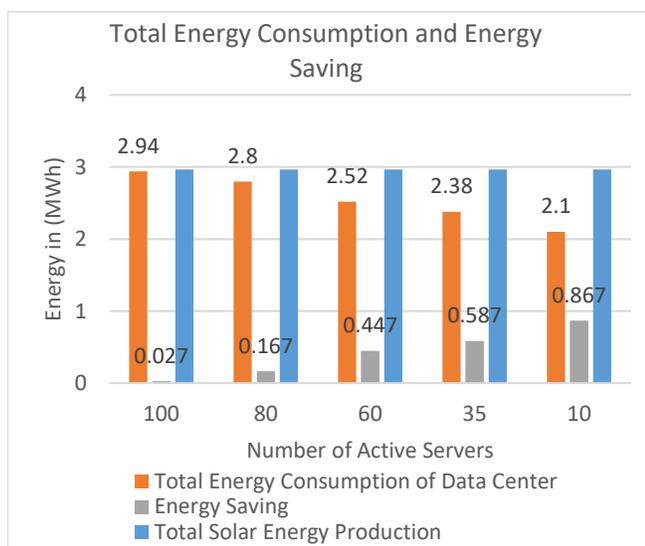

Fig. 9. Total Energy Consumption and Energy Saving

## V. CONCLUSION AND FUTURE WORKS

In this paper, we studied the potential of using renewable energy sources for data center implemented using of 100 traditional servers and other physical infrastructures and provided with 5 estimated batteries according to the energy provided by solar energy system. This paper is provided a model to analyze the renewable energy produced by 613 solar panel of size (6.5ft × 3.25ft) each contains 72 solar cell installed on an area of 1200m$^2$ in Mosul city to provide 2.967MWh for the total energy load of the data center. Because of the variation of the solar power production on a daily and seasonally basis, we propose an ON/OFF technology for energy efficient use of the solar power by switching OFF the idle servers in data center in case of there is a shortage in solar energy production especially during the night. The results show that to run only 4 batteries at night, the data center needs to switch OFF 65% until 90% of the total number of servers and consolidating the traffic to the 35% to 10% of the active servers to preserve power usage. On the other hand, solar power in a data center in addition to providing power and reducing the air pollution. It can increase the sustainability of power system, if the power is generated locally for a limited period, for example when power grid is not available, solar power increases the reliability of power system alongside the UPS and generators. The integration between different types power management technologies with different forms of renewable energy sources in the future will improve the sustainability and reduce the pollution of electricity production and enhance the energy efficiency.